\documentclass[10pt,prd,aps,onecolumn,preprintnumbers, showpacs,nofootinbib,superscriptaddress,notitlepage]{revtex4-1}

\usepackage{amsmath,graphicx,epsfig,amssymb}
\usepackage{inputenc}
\usepackage{ulem}
\usepackage{bigstrut}
\usepackage{slashed}
\usepackage{multirow}
\usepackage{subfigure}
\usepackage{dsfont}
\usepackage{xcolor}
\usepackage{soul}

\begin{document}

\title{Probing heavy meson lightcone distribution amplitudes  with heavy quark spin symmetry}
 
\author{Zhi-Fu Deng}
\affiliation{Shanghai Key Laboratory for Particle Physics and Cosmology, Key Laboratory for Particle Astrophysics and Cosmology (MOE), School of Physics and Astronomy, Shanghai Jiao Tong University, Shanghai 200240, P.R. China}

\affiliation{School of Science and Engineering, The Chinese University of Hong Kong, Shenzhen (CUHK-Shenzhen), Guangdong, 518172,  P.R. China}

\author{Wei Wang}  
\email{Corresponding author: wei.wang@sjtu.edu.cn}
\affiliation{Shanghai Key Laboratory for Particle Physics and Cosmology, Key Laboratory for Particle Astrophysics and Cosmology (MOE), School of Physics and Astronomy, Shanghai Jiao Tong University, Shanghai 200240, P.R. China}

\affiliation{Southern Center for Nuclear-Science Theory (SCNT), Institute of Modern Physics, Chinese Academy of Sciences, Huizhou 516000, Guangdong Province, P.R. China}

\author{Yan-Bing Wei}  
\email{Corresponding author: yanbing.wei@bjut.edu.cn}
\affiliation{School of Physics and Optoelectronic Engineering, Beijing University of Technology, Beijing
100124, P.R. China}

\author{Jun Zeng} 
\email{Corresponding author: zengj@sjtu.edu.cn}
\affiliation{Shanghai Key Laboratory for Particle Physics and Cosmology, Key Laboratory for Particle Astrophysics and Cosmology (MOE), School of Physics and Astronomy, Shanghai Jiao Tong University, Shanghai 200240, P.R. China}

\begin{abstract}
Building on a previous work~\cite{Han:2024min}, we illustrate  that the  leading-twist light-cone distribution amplitudes (LCDAs) defined in heavy-quark effective theory (HQET)  can be determined through lattice simulations of quasi-distribution amplitudes (quasi-DAs) with a large momentum component $P^z$.  Exploiting heavy-quark spin symmetry, we show that the LCDAs for a heavy pseudoscalar and vector meson in the context of HQET exhibit degeneracy, and the degeneracy allows for the utilization of quasi DAs for both pseudoscalar and vector mesons on the lattice.    We then derive the relevant short-distance  coefficients  for the matching between  LCDAs defined with QCD fields and HQET LCDAs at the one-loop level.  The incorporation of these three quasi DAs can not only confirm the methodology introduced in Ref.~\cite{Han:2024min} but also provides possible  insight into   power corrections.  Discrepancies between the corresponding results  offer a valuable perspective for estimating power corrections within the system which are imperative  for  precise investigations into heavy-meson LCDAs in the future particularly in the context of lattice QCD. 
\end{abstract}
\maketitle

\section{Introduction}

In hard exclusive decays of $B$ mesons like $B^0\to \pi^+\pi^-/K^+\pi^-$,  light-like distances are likely predominant. Under the   factorization schemes that can be established,  the heavy-meson light-cone distribution amplitudes (LCDAs)~\cite{Grozin:1996pq} play a vital role to calculate the decay amplitudes, as extensively discussed in the literature~\cite{Beneke:1999br,Beneke:2000ry,Keum:2000wi,Lu:2000em}.  The involved LCDAs are   defined as the hadron-to-vacuum matrix elements of light-ray operators  in heavy quark effective theory (HQET).  Understanding and modelling  these  LCDAs   (hereafter named as HQET LCDAs) are not only crucial within theoretical frameworks but also hold practical significance in predicting decay widths. These predictions, in turn, significantly contribute to determining Cabibbo-Kobayashi-Maskawa (CKM) matrix elements. Furthermore, a deep understanding of LCDAs is essential in the ongoing quest to explore new physics beyond the standard model, providing valuable insights into the potential existence and behavior of phenomena lying outside the  standard model.

Despite its significant importance, the heavy meson LCDAs have not been extensively  exploited  using first-principle tools such as lattice quantum chromodynamics (QCD). This challenge arises from the nature of heavy-meson LCDAs as light-like quantities that are defined in terms of the heavy-quark field within HQET. Due to this definition, heavy-meson LCDAs cannot be directly computed on the lattice, posing a major obstacle in their determination. Additionally, the presence of cusp divergences further complicates the situation~\cite{Braun:2003wx}, rendering the conventional moment expansion method, which has been successfully employed for light meson LCDAs, ineffective for heavy meson LCDAs. The combination of these factors presents a nearly formidable barrier to the calculation of heavy meson LCDAs using traditional lattice QCD techniques.


A recent study in Ref.~\cite{Han:2024min} addresses the obstacles mentioned above and presents an approach to overcoming them through a two-step matching technique. In the first step of this method, one can perform simulations to obtain the equal-time correlators, which serves as the basis for defining quasi distribution amplitudes (quasi DAs) characterizing a heavy meson at large momentum scales. By following the spirit of large momentum effective theory~\cite{Ji:2013dva,Ji:2014gla} (for recent reviews see Refs.~\cite{Cichy:2018mum,Ji:2020ect}),  in large momentum limit  $P^z\to \infty$ these quasi DAs can be expressed in terms of the LCDAs for a heavy meson defined within QCD, referred to as QCD LCDAs in subsequent discussions. By integrating out the scale associated with the heavy quark mass $m_Q$, it becomes possible to derive the LCDAs for a heavy meson in the context of HQET~\cite{Ishaq:2019dst,Zhao:2019elu,Beneke:2023nmj}, ultimately enabling a comprehensive understanding of the heavy-meson LCDAs through this innovative two-step matching procedure.  This approach eliminates the need for direct simulation of the HQET field on the lattice.

In this study, we adopt the methodology outlined in Ref.~\cite{Han:2024min}, emphasizing the utilization of heavy-quark spin symmetry to unlock additional avenues. Leveraging the robustness provided by heavy quark spin symmetry, we highlight the degeneracy of HQET LCDAs for heavy pseudoscalar and vector mesons. Consequently, three distinct types of equal-time correlators or quasi DAs emerge, facilitating the computation of HQET LCDAs. Through the utilization of these three matrix elements, we delve into the process of matching these distributions with both  QCD LCDAs and HQET LCDAs in a two-step manner. Our analysis extends to deriving the pertinent short-distance matching coefficients at the one-loop level, essential for advancing investigations into heavy-meson LCDAs. Notably, we observe that the matching kernels for the three matrix elements exhibit similarities, barring discrepancies in how they define decay constants. This observation serves a dual purpose: it not only validates the methodology proposed in Ref.~\cite{Han:2024min} but also sheds light on the potential for power corrections.  Consequently, disparities in the results obtained from the three   offer insights into estimating power corrections within the system, part of which are estimated in Ref.~\cite{Han:2024cht} in a dynamical way.

The remainder of this paper is structured as follows.  In Sec.~\ref{sec:HQET_LCDAs}, we present the definitions of HQET LCDAs. Subsequently, in Sec.~\ref{sec:matchingQCDLCDAsontoHQETLCDAs}, we delve into the process of matching QCD LCDAs onto HQET LCDAs. Within this section, we outline the computational steps, specifically deriving the matching kernel for scenarios where the momentum of the light quark corresponds to approximately $\Lambda_{\rm QCD}/m_H \times P^z$, commonly referred to as the peak region. The matching from quasi DAs to QCD LCDAs is given  in Sec.~\ref{sec:matchingquasiDAsontoQCDLCDAs}. A brief  summary is provided in the concluding section. Additional details  are included in the appendix.

\section{Heavy meson LCDAs with heavy quark spin symmetry}
\label{sec:HQET_LCDAs}

The HQET Lagrangian in heavy-quark limit ~\cite{Georgi:1990um,Grinstein:1990mj} reads
\begin{eqnarray}
\mathcal{L}_{\rm HQET} = \bar{h}_v(x) i v \cdot D h_v(x). 
\end{eqnarray} 
Here $h_v$ denotes the effective heavy quark field with four-velocity $v$ defined as
\begin{eqnarray}
    h_v(x) = e^{im_Q v \cdot x} \frac{1+\slashed{v}}{2}Q(x),
\end{eqnarray}
while $Q(x)$ is the heavy quark field in full QCD. It should be noticed that in the heavy-quark limit the interaction is spin independent, which is known as heavy-quark spin symmetry. Under this symmetry, the heavy pseudoscalar and vector  mesons belong to the same multi-plet. As a result,  the decay constants for the pseudoscalar meson $H$ and the vector $H^*$ are the same.

The heavy-quark spin symmetry implies the leading-twist LCDAs for $H$ and $H^*$  are also the same.  In HQET, the  leading-twist  LCDAs of  a heavy pseudoscalar and vector mesons  are defined  as~\cite{Grozin:1996pq}: 
\begin{eqnarray}\label{eq:HQETLCDA}
    \langle H(p_H)|O_{v}^P(tn_+) |0\rangle = -i \tilde f_H   m_H n_+\cdot v \int_0^\infty d\omega e^{i \omega t n_+\cdot v} \varphi_+(\omega;\mu),\nonumber\\
    \langle H^*(p_H,\eta)| O_{v}^{||}(tn_+) |0\rangle =\tilde  f_H  m_H n_+ \cdot \eta^{*} \int_0^\infty d\omega e^{i \omega t n_+\cdot v} \varphi_+(\omega;\mu),\nonumber\\
    \langle H^*(p_H,\eta)| O_{v}^{\perp\mu}(tn_+) |0\rangle =\tilde  f_H  m_H n_+\cdot v \eta_{\perp}^{*\mu}\int_0^\infty d\omega e^{i \omega t n_+\cdot v} \varphi_+(\omega;\mu),
\end{eqnarray} 
where $\tilde f_H$ is the decay constant defined in  HQET, and the involved  operators are given as:  
\begin{eqnarray}
O_{v}^P(tn_+) &=& \bar{h}_v(0) \slashed{n}_+\gamma_5[0,tn_+]q_s(tn_+),  \nonumber\\
O_{v}^{||}(tn_+) &=& \bar{h}_v(0) \slashed{n}_+[0,tn_+]q_s(tn_+), \nonumber\\
 O_{v}^{\perp\mu}(tn_+)&=& \bar{h}_v(0) \slashed{n}_+ \gamma^\mu_{\perp}[0,tn_+]q_s(tn_+),
\end{eqnarray}
where $[0,tn_+]$ is a finite-distance Wilson line with $n_+$ is a light-like vector.  The $q_s$ represents a light and soft quark field. The $[0,tn_+]$ denotes the Wilson line which connects the soft quark field $q_s$ and the $h_v$ and ensures the gauge invariance. The hardron state $|H^{(*)}(p_H)\rangle$ is related to the HQET state $|H_v\rangle$ by $|H^{(*)}(p_H)\rangle=\sqrt{m_H}\big[|H_v\rangle  + \mathcal{O}(1/m_H)\big]$, with $m_H$ being the hadron mass. At leading power the masses for both pseudoscalar and vector mesons are  equal to the heavy quark mass $m_Q$. 

The HQET LCDAs is related to the QCD LCDAs of heavy mesons~\cite{Ishaq:2019dst,Zhao:2019elu,Beneke:2023nmj}. 
To obtain the relations, it is convenient to work in a frame in which the heavy meson is fast moving.
Ref.~\cite{Beneke:2023nmj} points out that HQET LCDAs is boost invariant and can also be defined in boosted HQET (bHQET)~\cite{Fleming:2007qr,Fleming:2007xt}. 
We use  the bHQET field $h_n$ as \cite{Dai:2021mxb}
\begin{eqnarray}
\label{eq:HQETfield}
h_n(x)\equiv \sqrt{\frac{2}{n_{+}\cdot v}}e^{i m_Q v \cdot x}\frac{\slashed{n}_-\slashed{n}_+}{4}Q(x)\,,
\end{eqnarray}
which projects $Q(x)$ onto its large component $\frac{\slashed{n}_- \slashed{n}_+}{4}Q(x)$, and then subtracts the rapid phase variations from the large momentum piece $m_Q v$  through the exponential factor as in HQET, leaving the residual momentum. The normalization factor in Eq.~\eqref{eq:HQETfield} is chosen such that the $h_n$ field will have a scaling independent of the boost as shown below. Therefore, the definitions of  LCDAs of heavy meson matrix elements in bHQET are
\begin{eqnarray}
   \langle H(p_H)|\mathcal{O}_b^P(\omega) |0\rangle &=& -i \tilde f_H    \varphi_+(\omega;\mu),\nonumber\\
\langle H^*(p_H,\eta)| \mathcal{O}_b^{||}(\omega)|0\rangle &=&\tilde f_H \frac{n_+ \cdot \eta^{*}}{n_+ \cdot v} \varphi_+(\omega;\mu),\nonumber\\
\langle H^*(p_H,\eta)| \mathcal{O}_b^{\perp \mu}(\omega) |0\rangle &=& \tilde f_H \eta_{\perp}^{*\mu} \varphi_+(\omega;\mu). 
\end{eqnarray} 
The involved  bHQET operators are constructed as: 
\begin{eqnarray}
\mathcal{O}_b^P(\omega)&=&\frac{1}{m_H}\int\frac{dt}{2\pi}e^{-it\omega n_+\cdot v}\sqrt{\frac{n_+\cdot v}{2}}\bar{h}_n(0) \slashed{n}_+\gamma_5[0,tn_+]\xi_{sc}(tn_+), \nonumber\\
\mathcal{O}_b^{||}(\omega)&=&\frac{1}{m_H}\int\frac{dt}{2\pi}e^{-it\omega n_+\cdot v}\sqrt{\frac{n_+\cdot v}{2}}\bar{h}_n(0) \slashed{n}_+[0,tn_+]\xi_{sc}(tn_+), \nonumber\\
\mathcal{O}_b^{\perp \mu}(\omega)&=&\frac{1}{m_H}\int\frac{dt}{2\pi}e^{-it\omega n_+\cdot v}\sqrt{\frac{n_+\cdot v}{2}}\bar{h}_n(0) \slashed{n}_+\gamma^\mu_\perp[0,tn_+]\xi_{sc}(tn_+),
\end{eqnarray}
where the soft-collinear field $\xi_{sc}=\frac{\slashed{n}_- \slashed{n}_+}{4}q_{sc}(x)$ describes the light anti-quark in the heavy meson in the boosted frame.

\section{QCD LCDAs and Matching onto HQET LCDAs}
\label{sec:matchingQCDLCDAsontoHQETLCDAs}

The non-perturbative HQET LCDAs defined in the previous section includes the degree-of-freedom below the $\Lambda_{\rm QCD}$ scale. The QCD LCDAs corresponds to physics below the $m_Q$ scale. In this section, we give the definitions of QCD LCDAs under the  soft-collinear effective theory (SCET)~\cite{Bauer:2000yr,Bauer:2001yt,Beneke:2002ph,Beneke:2002ni,Becher:2014oda}  and factorize the QCD LCDAs into a convolution of the $m_Q$-scale jet function with the HQET LCDAs.

To derive the factorization formula, we work in a frame in which the heavy meson is energetic. 
Thus we describe the light quark $q(x)$ and heavy quark $Q(x)$ fields employing SCET framework 
\begin{align}
q(x) =&~ \xi_C(x) +\eta_C(x),
&
\xi_C(x) =&~ \frac{\slashed{n}_-\slashed{n}_+}{4} q(x),
&
\eta_C(x) =&~ \frac{\slashed{n}_+\slashed{n}_-}{4} q(x), 
\nonumber \\
Q(x) =&~ \xi^{(Q)}_C(x) +\eta^{(Q)}_C(x),
&
\xi^{(Q)}_C(x) =&~ \frac{\slashed{n}_-\slashed{n}_+}{4} Q(x),
&
\eta^{(Q)}_C(x) =&~ \frac{\slashed{n}_+\slashed{n}_-}{4} Q(x). 
\end{align}
The SCET operators defining the QCD LCDAs are
\begin{align}
\mathcal{O}^{P}_C(u)
=&\int\frac{dt}{2\pi}e^{-iutn_+\cdot p}\bar{\xi}^{(Q)}_C(0)\slashed n_+ \gamma_5 
\left[0, t n_{+}\right] \xi_C (tn_+)
= \int\frac{dt}{2\pi}e^{-iutn_+\cdot p}\bar Q(0)\slashed n_+ \gamma_5 
\left[0, t n_{+}\right] q (tn_+),
\nonumber \\
\mathcal{O}^{\|}_C(u)
=&\int\frac{dt}{2\pi}e^{-iutn_+\cdot p}\bar{\xi}^{(Q)}_C(0)\slashed n_+ 
\left[0, t n_{+}\right] \xi_C (tn_+)
= \int\frac{dt}{2\pi}e^{-iutn_+\cdot p}\bar Q(0)\slashed n_+ 
\left[0, t n_{+}\right] q(tn_+),
\nonumber \\
\mathcal{O}^{\perp\mu}_C(u)
=&\int\frac{dt}{2\pi}e^{-iutn_+\cdot p}\bar{\xi}^{(Q)}_C(0) \slashed{n}_+ \gamma^\mu_\perp
\left[0, t n_{+}\right] \xi_C (tn_+)
= \int\frac{dt}{2\pi}e^{-iutn_+\cdot p}\bar Q(0) \slashed{n}_+ \gamma^\mu_\perp
\left[0, t n_{+}\right] q(tn_+),
\end{align}
where $u$ stands for the fraction of momentum of the light quark compared to the momentum of the meson. 
Once taking the matrix element of the SCET operator, the momentum $p$ will become the momentum of the heavy meson $p_H$. 
Employing the equation of motions $\slashed{n}_+\eta_C=0$ and $\bar \eta^{(Q)}_C \slashed{n}_+=0$, one obtains that the three SCET operators are the same to the three QCD operators, respectively.

The QCD LCDAs of heavy meson are defined in terms of SCET operators as~\cite{Braun:1988qv,Ball:1998sk,Hardmeier:2003ig}
\begin{align}
\langle H\left(p_H\right)|\mathcal{O}^{P}_C(u)| 0\rangle
=&~ -i f_P \phi_{P}(u),  
\nonumber \\
\langle H^*\left(p_H,\eta\right)|\mathcal{O}^{\|}_C(u)| 0\rangle
=&~ f_\| \frac{m_H}{n_+\cdot p_H} n_+\cdot \eta^\ast \phi_{\|}(u),
\nonumber \\
\langle H^*\left(p_H,\eta\right)|\mathcal{O}^{\perp\mu}_C(u)| 0\rangle
=&~  f_\perp(\mu) \eta_{\perp}^{\ast\mu} \phi_{\perp}(u).
\end{align}
The relationship of the HQET decay constants and the QCD decay constants are~\cite{Eichten:1989zv,Ji:1991pr,Broadhurst:1994se,Manohar:2000dt}
\begin{eqnarray}
f_P &=&\tilde f_H(\mu) \Big[1- \frac{\alpha_s C_F}{4\pi} 
\Big( 3\ln\frac{\mu}{m_Q} +2 \Big) +{\cal O}(\alpha^2_s)\Big] ,
\nonumber \\
f_{||} &=& \tilde f_H(\mu) \Big[1- \frac{\alpha_s C_F}{4\pi} 
\Big( 3\ln\frac{\mu}{m_Q} +4 \Big) +{\cal O}(\alpha^2_s)\Big],
\nonumber \\
f_\perp(\mu) &=& \tilde f_H(\mu) \Big[1- \frac{\alpha_s C_F}{4\pi} 
\Big( 5\ln\frac{\mu}{m_Q} +4 \Big) +{\cal O}(\alpha^2_s)\Big]. 
\label{eq:decayconstant}
\end{eqnarray}
It should be noticed that  $f_P$ and $f_{||}$ are scale-independent, while we did not distinguish the renormalization scales in $f_\perp(\mu) $ and $\tilde f_H(\mu)$. At the scale $m_Q$, $f_\perp(m_Q)=f_{||}$.

It is clear that QCD LCDAs receive both a short-distance contribution from the $m_Q$ scale and a long-distance contribution from the $\Lambda_{\rm QCD}$ scale.
The two separated scales generate large logarithms of the form $\ln m_Q/\Lambda_{\rm QCD}$, which should be resummed properly with the renormalization group (RG) eqnarray technique.
This requires that the LCDA be factorized into a convolution of a $m_Q$ scale jet function with heavy-meson bHQET LCDA~\cite{Beneke:2023nmj}.  To obtain the factorization formula, we follow Ref.~\cite{Beneke:2023nmj} and distinguish the momentum fraction $u$ to be small and large and call the two regions the peak region and the tail region, respectively. 
At the tail region, where $u\sim {\cal O}(1)$, the QCD LCDA is not related the form of the bHQET LCDA, and thus we will not consider the tail region in this paper.

At the peak region, it is conjectured that the QCD LCDA is factorized as
\begin{align}
\phi_i(u)= {\displaystyle \frac{\tilde f_H}{f_i}} 
\mathcal{J}^{i}(u,\omega) \otimes \varphi_+(\omega),
\qquad 
u\sim {\cal O}(\frac{\Lambda_{\rm QCD}}{m_Q}). 
\end{align}
With the factorization formula at hand, the large logarithms could be resummed. The peak-region jet function satisfies the form
\begin{eqnarray}
\mathcal{J}^{i}(u,\omega) = 
\theta(m_H-\omega) \delta\bigg(u - \frac{\omega}{m_H}\bigg)  \mathcal{J}^{i}_{\rm peak}(m_H) \,,
\qquad i=P,~\|,~\perp,
\end{eqnarray}
where $\mathcal{J}^{i}_{\rm peak}$ depends only on $m_Q$ and $m_H$ since $u m_H$ and $\omega$ are nonperturbative scale  variables. The $\delta$ function comes from the momentum conservation of the light quark in the heavy meson.
Then one derives the QCD LCDAs at the peak region
\begin{align}
\phi_{i}(u) 
=&~ \frac{\tilde f_H}{f_i} m_H {\cal J}^i_{\rm peak}(m_H) \varphi_+(u m_H). 
\end{align}
The heavy-meson QCD LCDAs defined by different Dirac structures at the peak region are related to the single bHQET LCDA at leading power~\cite{Beneke:2023nmj,Han:2024min}.

\subsection{The calculation procedure of the jet function}

\begin{figure}
\centering
\includegraphics[width=0.62\textwidth]{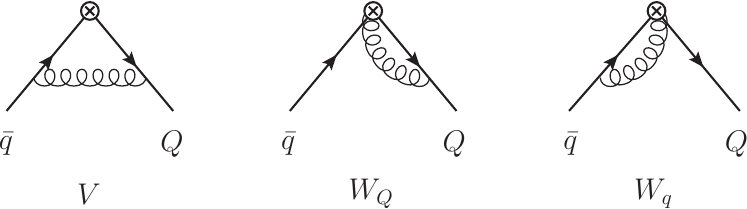}
\caption{\small The one-loop diagrams relevant for calculating the jet function.}
\label{fig:scet}
\end{figure}

For a general jet function $\mathcal{J}^{i}$, which is insensitive to the physics below the $\Lambda_{\rm QCD}$ scale, it could be determined by the matching relation with on-shell external states
\begin{align}
\langle Q(p_Q)\bar{q}(p_q)|\mathcal{O}^{i}_C(u)|0\rangle
=\int^\infty_0 d\omega \mathcal{J}^{i}(u,\omega) 
\langle Q(p_Q)\bar{q}(p_q)|\mathcal{O}^{i}_b(\omega)|0\rangle.
\end{align}
We use the condition $p^\alpha_Q=m_Q v^\alpha$, $p^\alpha_q=n_+p_q n^\alpha_-$.
By calculating the one-loop correction to the matrix elements of the SCET operator and the bHQET operator, one can derive the jet function at NLO.
The matrix elements of the operators are
\begin{align}
\langle Q(p_Q)\bar{q}(p_q)|\mathcal{O}^{i}_C(u)|0\rangle
=\frac{1}{n_+p_H} \bar u(p_Q) \Gamma v(p_q) 
\Big[ \delta(u-s) + \frac{\alpha_s C_F}{4\pi} M^{(1)}_{i}(u,s)
+{\cal O}(\alpha_s^2)\Big],
\nonumber \\
\langle Q(p_Q)\bar{q}(p_q)|\mathcal{O}^{i}_b(\omega)|0\rangle
=\frac{1}{n_+p_H} \bar u(p_Q) \Gamma v(p_q) 
\Big[ \delta(\omega-\omega_0) 
+ \frac{\alpha_s C_F}{4\pi} N^{(1)}_{i}(\omega,\omega_0)
+{\cal O}(\alpha_s^2)\Big] ,
\end{align}
where $s=n_+ p_q/n_+ (p_q+p_Q)\equiv n_+p_q/n_+ p_H$ and $\omega_0=n_+ p_q/n_+v$. 

The three diagrams in Fig.~\ref{fig:scet} will contribute both to the SCET side and the bHQET side at NLO. 
The renormalized amplitudes are
\begin{align}
M^{(1)}_{i}(u,s) =&~ V_{i}(u,s) + W_{Q,i}(u,s) +W_{q,i}(u,s)
+ \frac{1}{2} Z^{OS(1)}_Q \delta(u-s) + Z^{(1)}_{{\cal O}^{i}_C}(u,s),
\nonumber \\
N^{(1)}_{i}(\omega,\omega_0) =&~ V^{\rm bHQET}_{i}(\omega,\omega_0) 
+ W^{\rm bHQET}_{Q,i}(\omega,\omega_0) 
+W^{\rm bHQET}_{q,i}(\omega,\omega_0)
+ Z^{(1)}_{{\cal O}^{i}_b}(\omega,\omega_0).
\end{align}
From the matching relation, the jet function at NLO is determined as
\begin{align}
{\cal J}^{i,(0)}(u,\omega) 
=&~ \delta(\frac{\omega}{m_H} - u) \theta(m_H-\omega),
\nonumber \\
{\cal J}^{i,(1)}(u,\omega) 
=&~ \theta(m_H-\omega) \Big[ M^{(1)}_{i}(u,\frac{\omega}{m_H})
-m_H N^{(1)}_{i}(um_H,\omega)
\Big] .
\end{align}
We will regulate both the UV and IR divergence with dimensional regulation scheme.
Then the $W_q$ diagram is scaleless and generates $W_{q,i}=0$ and $W^{\rm bHQET}_{q,i}=0$.

\subsection{The jet function at NLO with the method of regions}

The method of regions, which is suitable for computing diagrams with multi scales~\cite{Beneke:1997zp,Smirnov:2002pj}, is employed to compute the SCET diagrams.
The hard-collinear region and soft-collinear region will contribute to the SCET side.
Separating the two regions, the jet function is
\begin{align}
{\cal J}^{i,(1)}(u,\omega) 
=&~ \theta(m_H-\omega) \bigg\{ 
W^{hc}_{Q,i}(u,\frac{\omega}{m_H})
+ \frac{1}{2} Z^{OS(1)}_Q \delta(u-\frac{\omega}{m_H}) 
+ Z^{(1)}_{{\cal O}^{i}_C}(u,s)
-m_H Z^{(1)}_{{\cal O}^{i}_b}(um_H,\omega)
\nonumber \\
&~
+\Big[V^{sc}_{i}(u,\frac{\omega}{m_H})
-m_H V^{\rm bHQET}_{i}(um_H,\omega)\Big] +
 \Big[  W^{sc}_{Q,i}(u,\frac{\omega}{m_H})
-m_H W^{\rm bHQET}_{Q,i}(um_H,\omega) \Big]
\bigg\}.
\label{eq:jet.match}
\end{align}
The diagram $V$ only receive soft-collinear region contribution since the hard-collinear region breaks the momentum conservation.
The soft-collinear region contribution of SCET is diagram by diagram the same as the bHQET results.
We confirm that 
\begin{align}
V^{sc}_{i}(u,\frac{\omega}{m_H}) = m_H V^{\rm bHQET}_{i}(um_H,\omega),
\qquad
W^{sc}_{Q,i}(u,\frac{\omega}{m_H}) =
m_H W^{\rm bHQET}_{Q,i}(um_H,\omega),
\end{align}
and see App.~\ref{app:jet} for details.
Thus the second line in Eq.~(\ref{eq:jet.match}) cancels exactly, we are left to calculate the terms in the first line.

For a general SCET operator the amplitudes of the $W_Q$ diagram is parameterized as
\begin{align}
&~ A_{W_Q,i} = \frac{\alpha_s C_F}{4\pi} \bar \xi^Q_C 
\Big[ f_1(p_Q,p_q) + f_2(p_Q,p_q) \frac{\slashed n_+ \slashed n_-}{4}\Big] \Gamma_i  \xi_C,
\label{eq:WQ}
\end{align}
where the form factors $f_1$ and $f_2$ depend on $p^2_Q$ and $p_Q\cdot p_q$. 
We have employed $p_{q\perp}=0$ which means that the independent Dirac structure of the form factors could only be $1$, $\slashed n_+$ and $\slashed n_+ \slashed n_-$. 
Recall that $\Gamma_i$ in the eqnarray is $\slashed n_+ \gamma_5$, $\slashed n_+$ and $\slashed n_+ \gamma^\mu_\perp$, we are left with the two form factors.
Then one directly derives 
\begin{align}
W^{hc}_{Q,P} =W^{hc}_{Q,\|} =W^{hc}_{Q,\perp} = 
\frac{\alpha_s C_F}{4\pi}
\Big[ f_1(p_Q,p_q) + f_2(p_Q,p_q)\Big] .
\end{align}

Now we need to calculate the renormalization constants of the SCET operators and the bHQET operators.
For the SCET operators, the $W_Q$ diagram generates the same UV divergence as one could see from Eq.~(\ref{eq:WQ}).
The amplitudes of the $W_q$ diagram of a SCET operator can be parameterized as
\begin{align}
&~ A_{W_q,i} = \frac{\alpha_s C_F}{4\pi} \bar \xi^Q_C \Gamma_i
\Big[ g_1(p_Q,p_q) + g_2(p_Q,p_q) \frac{\slashed n_- \slashed n_+}{4}\Big]  \xi_C,
\end{align}
which indicates that the UV divergences of the three SCET operators form the $W_q$ are the same.
The amplitude of the diagram $V$, which can not be parameterized as that for $W_Q$ and $W_q$, is calculated directly.
At leading power the UV divergences are
\begin{eqnarray}
V^{\rm UV}_P =V^{\rm UV}_\| = V^{\rm UV}_\perp 
= 0.
\end{eqnarray}
It is worth to mention that at subleading power, the UV divergences from the V diagram are different for different SCET operators.
The renormalization constants of the three SCET operators $ Z^{(1)}_{{\cal O}^{i}_C}$ are identical.

For the bHQET operators, the $W_Q$ and $W_q$ diagrams introduce
\begin{align}
W^{\rm bHQET}_{Q,P} =&~ W^{\rm bHQET}_{Q,\|} = W^{\rm bHQET}_{Q,\perp} ,
\nonumber \\
W^{\rm bHQET}_{q,P} =&~ W^{\rm bHQET}_{q,\|} = W^{\rm bHQET}_{q,\perp}.
\end{align}
The above eqnarrays are derived in the same way as for the SCET case.
For the $V$ diagram, we could also parameterized it as
\begin{align}
A^{\rm bHQET}_{V,i} =&~ \frac{\alpha_s C_F}{4\pi} \bar h_n \Gamma_i
\Big[ d_1(p_q) + d_2(p_q) \frac{\slashed n_- \slashed n_+}{4}\Big]  \xi_{sc}.
\end{align}
Thus the renormalization constants of the three bHQET operators $ Z^{(1)}_{{\cal O}^{i}_h}$ are the same.
Then the jet functions are
\begin{align}
\label{eq:jet}
\mathcal{J}^i(u,\omega) =
\theta(m_H-\omega) \delta(u - \frac{\omega}{m_H}) 
\bigg( 1 + \frac{\alpha_s C_F}{4\pi} 
\mathcal{J}^{(1)}_{\rm peak}(m_H) + \mathcal{O}(\alpha_s^2)\bigg) \,,
\qquad
i=P,~\|,~\perp,
\end{align}
with~\cite{Beneke:2023nmj}
\begin{align}
\mathcal{J}^{(1)}_{\rm peak}(m_H) = \frac{1}{2} \ln^2\frac{\mu^2}{m^2_H}
+\frac{1}{2} \ln\frac{\mu^2}{m^2_H} +\frac{\pi^2}{12}+2.
\end{align}

\section{Quasi DAs to QCD LCDAs}
\label{sec:matchingquasiDAsontoQCDLCDAs}

In LaMET~\cite{Ji:2013dva,Ji:2014gla} (for recent reviews see Refs.~\cite{Cichy:2018mum,Ji:2020ect}), one can make use of equal-time correlation functions which are named as the quasi-DAs to extract the QCD LCDAs. In this framework, the unit vector of the $z$ direction is denoted by $n^{\mu}_z=(0,0,0,-1)$. We introduce the non-local bilinear operators, in which the fermion fields are  separated on the $z$ direction: 
\begin{eqnarray}
 {\mathcal{\tilde O}}_i(x)=\int \frac{dz n_z\cdot P}{2\pi}e^{-i x z n_z\cdot P}\bar Q(0)\Gamma_i [0,zn_z]q(zn_z),
\end{eqnarray}
where $\Gamma_i=\gamma^{z/t}\gamma_5, \gamma^{z/t},  \gamma^{z/t}\gamma^{\mu}_{\perp}$ for $i=P,\parallel,\perp$, respectively. Then, the quasi-DAs of the transverse and longitudinal components of a vector meson are defined by the matrix elements of the operators as
\begin{eqnarray}
\langle H\left(p_H\right)| {\mathcal{\tilde O}}_P(x)| 0\rangle
&=&-i \tilde f_P P^{z/t} \tilde \phi_{P}(x),  \nonumber\\
\langle H^\ast\left(p_H,\eta\right)| {\mathcal{\tilde O}}_{||}(x)| 0\rangle
&=&\tilde f_{||}m_{H^*} \eta^{*, z/t} \tilde\phi_{||}(x),  \nonumber\\
\langle H^\ast\left(p_H,\eta\right)| {\mathcal{\tilde O}}^\mu_{\perp}(x)| 0\rangle
&=&\tilde f_{\perp} P^{z/t}   \eta_{\perp}^{*\mu} \tilde\phi_{\perp}(x). 
\end{eqnarray}

The quasi DAs can be factorized as: 
\begin{eqnarray}
\tilde \phi_i(x) = \int dy \Bigg[\bigg(\delta(x-y)+C_i^{(1)}(x,y,\mu) -C_{CT}^{(1)}\bigg)\phi_i(y,\mu)+\mathcal{O}\Big(\frac{\Lambda_{QCD}^2}{{(yP^z)}^2},\frac{m_H^2}{{(P^z)}^2}\Big)\Bigg]. 
\end{eqnarray}
The matching kernels $C_i(x,y,\mu)$ for these quasi DAs are given in Refs.~\cite{Xu:2018mpf,Liu:2018tox},

\begin{align}
&C^{(1)}_i\left(\Gamma,x,y,\frac{P_z}{\mu}\right)=\frac{\alpha_s C_F}{2\pi}\left\{
\begin{array}{lc}
\left[H_1(\Gamma,x,y)\right]_{+(y)}	        	& x<0<y\\
\left[H_2(\Gamma,x,y,P^z/\mu)\right]_{+(y)}		& 0<x<y\\
\left[H_2(\Gamma,1-x,1-y,P^z/\mu)\right]_{+(y)}	& y<x<1\\
\left[H_1(\Gamma,1-x,1-y)\right]_{+(y)}			& y<1<x
\end{array}\right., \\
&C_{CT}^{(1)}\left(\Gamma,x,y\right)=\left\{
\begin{array}{lc}
\displaystyle \frac{3\alpha_s C_F}{4\pi}\left|\frac{1}{x-y}\right|_+    & \Gamma=\gamma^z\gamma_5{\;\rm and\;}\gamma^t\\ 
\displaystyle \frac{\alpha_s C_F}{\pi}\left|\frac{1}{x-y}\right|_+	    & \Gamma=\gamma^z\gamma_\perp
\end{array}\right., 
\end{align}
where
\begin{align}
H_1(\Gamma,x,y)&=\left\{
\begin{array}{ll}
\frac{1+x-y}{y-x}\frac{1-x}{1-y}\ln\frac{y-x}{1-x}+\frac{1+y-x}{y-x}\frac{x}{y}\ln\frac{y-x}{-x} & \Gamma=\gamma^z\gamma_5{\;\rm and\;}\gamma^t\\
\frac{1}{y-x}\frac{1-x}{1-y}\ln\frac{y-x}{1-x}+\frac{1}{y-x}\frac{x}{y}\ln\frac{y-x}{-x} & \Gamma=\gamma^z\gamma_\perp
\end{array} \right.\, ,\\
H_2\left(\Gamma,x,y,\frac{P_z}{\mu}\right)&=\left\{
\begin{array}{ll}
\frac{1+y-x}{y-x}\frac{x}{y}\ln\frac{4x(y-x)(P^z)^2}{\mu^2}+\frac{1+x-y}{y-x}\left(\frac{1-x}{1-y}\ln\frac{y-x}{1-x}-\frac{x}{y}\right) & \Gamma=\gamma^z\gamma_5\\
\frac{1+y-x}{y-x}\frac{x}{y}\left(\ln\frac{4x(y-x)(P^z)^2}{\mu^2}-1\right)+\frac{1+x-y}{y-x}\frac{1-x}{1-y}\ln\frac{y-x}{1-x} & \Gamma=\gamma^t\\
\frac{1}{y-x}\frac{x}{y}\ln\frac{4x(y-x)(P^z)^2}{\mu^2}+\frac{1}{y-x}\left(\frac{1-x}{1-y}\ln\frac{y-x}{1-x}-\frac{x}{y}\right) & \Gamma=\gamma^z\gamma_\perp
\end{array} \right. \, .
\end{align}
These results have been used in the calculation of light meson LCDAs from lattice QCD~\cite{Zhang:2017bzy,Zhang:2017zfe,Zhang:2020gaj,Hua:2020gnw,LatticeParton:2022zqc,Gao:2022vyh,Holligan:2024umc,Baker:2024zcd}, where it should be noticed that the long-distance behavior should be properly handled with the hybrid renormalization scheme~\cite{Ji:2020brr}.

\section{Conclusion}
The determination of heavy-meson LCDAs is crucial for understanding heavy meson decays, and a recent analysis offers a pioneering approach to precisely determine the heavy meson  leading twist  LCDAs from first principles~\cite{Han:2024min}, in which it is pointed out that  the determination of HQET LCDAs can be accomplished by engaging in lattice simulations of quasi DAs that involve a significant momentum component $P^z$. This proposal entails a step-by-step integration process that effectively removes $P^z$ and $m_H$, thereby disentangling various dynamic scales. By integrating out $P^z$, we can establish a link between quasi-DAs and QCD LCDAs, while the elimination of the $m_H$ scale enables a connection between QCD LCDAs and HQET LCDAs.  

In this work, we have pointed out that as a result of heavy-quark spin symmetry, the LCDAs at for a heavy pseudoscalar and vector meson within the framework of  HQET become indistinguishable. Thus one can make use of three different equal time matrix elements and determine the same HQET LCDA. However during process, the short-distance contributions are different. To accommodate these differences,  we have derived pertinent short-distance matching coefficients at the one-loop level, which are instrumental for forthcoming investigations into the LCDAs of heavy mesons. Consistencies  and discrepancies between the corresponding results  offer a valuable perspective for estimating power corrections within the system which are imperative  for  precise investigations into heavy-meson LCDAs in the future particularly in the context of lattice QCD.

Generalizations of this work can include the calculation of the QCD LCDAs for a vector heavy meson at the tail region following Ref.~\cite{Beneke:2023nmj}, which allows to obtain a full shape for QCD LCDAs by merging the results in the peak region. In addition the impact of heavy quark spin symmetry in the approach to construct the quasi DAs with HQET field as in Ref.~\cite{Kawamura:2018gqz,Wang:2019msf,Xu:2022guw,Hu:2023bba} deserves further explorations. 




\section*{Acknowledgement}

We thank Chao Han, Yu-Ji Shi,  Yu-Ming Wang, Ji Xu,  Jia-Lu Zhang,  Qi-An Zhang, and Shuai Zhao for fruitful discussions.  This work is supported in part by Natural Science Foundation of China under grant No. 12335003, 12305099, 12205180, 12125503, and 12147140. J.Z. is also partially supported by the Project funded by China Postdoctoral Science Foundation under Grant No. 2022M712088.

\appendix

\section{Jet function}
\label{app:jet}

The bHQET diagrams reproduce the soft-collinear region result of the SCET side.
For the diagram $V$, the SCET amplitude and the bHQET amplitude are
\begin{eqnarray}
V^{sc}_{i}(u,s) &=& -i g^2_s C_F \frac{1}{n_+\cdot p_H} 
\int \frac{d^D l}{(2\pi)^D} \bar u(p_Q)
\frac{1}{n_+\cdot v v\cdot l} \frac{\slashed n_+\slashed n_-}{4} 
\delta(u +\frac{n_+\cdot l}{n_+\cdot p_H}-s) \Gamma 
\frac{n_+\cdot l-s n_+\cdot p_H}{(l-p_q)^2 l^2}\frac{\slashed n_-\slashed n_+}{4}
v(p_q) ,
\nonumber \\
V^{\rm bHQET}_{i}(\omega,\omega_0) &=& -i g^2_s C_F \frac{1}{n_+\cdot p_H} 
\int \frac{d^D l}{(2\pi)^D} \bar u(p_Q)
\frac{n_-\cdot v}{v\cdot l} \frac{\slashed n_+\slashed n_-}{4} 
\delta(\omega +\frac{n_+\cdot l}{n_+\cdot v}-\omega_0) \Gamma 
\frac{n_+\cdot (l-\omega_0 v)}{(l-p_q)^2l^2}\frac{\slashed n_-\slashed n_+}{4}
v(p_q).
\end{eqnarray}
For the diagram $W_Q$, the SCET amplitude and the bHQET amplitude are
\begin{eqnarray}
W^{sc}_{Q,i}(u,s) &=& i
g^2_s C_F \frac{1}{n_+\cdot p_H} 
\int \frac{d^D l}{(2\pi)^D} \bar u(p_Q)
\frac{n_+\cdot v}{v\cdot l n_+\cdot l l^2} \frac{\slashed n_+\slashed n_-}{4} 
\Gamma 
\Big[\delta(u +\frac{n_+\cdot l}{n_+\cdot p_H}-s)
-\delta(u -s)\Big]
v(p_q) ,
\nonumber \\
W^{\rm bHQET}_{Q,i}(\omega,\omega_0) &=& i
g^2_s C_F \frac{1}{n_+\cdot p_H} 
\int \frac{d^D l}{(2\pi)^D} \bar u(p_Q)
\frac{n_+\cdot v}{v\cdot l n_+\cdot l l^2} \frac{\slashed n_+\slashed n_-}{4} 
\Gamma 
\Big[\delta(\omega +\frac{n_+\cdot l}{n_+\cdot v}-\omega_0)
-\delta(\omega -\omega_0)\Big]
v(p_q).
\end{eqnarray}

The hard-collinear region reault for the $W_Q$ diagram is
\begin{align}
W_{Q,i}^{hc}=&~ \left(\frac{\mu^2}{m_Q^2}\right)^\epsilon 2 e^{\epsilon \gamma_E} \Gamma(\epsilon) \frac{\delta(u-s)}{2\epsilon(1-2\epsilon)}.
\end{align}
For the completeness of the paper, we also collect here the renormalization constants necessery for deriving the jet function
\begin{align}
Z^{OS(1)}_Q=&~ 
-\frac{3}{\epsilon}-3\ln\frac{\mu^2}{m_Q^2}-4,
\nonumber \\
Z^{(1)}_{{\cal O}^{i}_C}(u,s)=&~
\frac{2}{\epsilon}\left[ \frac{\bar u}{\bar s} \frac{\theta(u-s)}{u-s}+\frac{u}{s}\frac{\theta(s-u)}{s-u}  \right]_{s+} ,
\nonumber \\
Z^{(1)}_{{\cal O}^{i}_b}(\omega,\omega_0) =&~
-\frac{2\omega}{\epsilon}\left[ \frac{\theta(\omega-\omega_0)}{\omega(\omega-\omega_0)}+\frac{1}{\omega_0}\frac{\theta(\omega_0-\omega)}{\omega_0-\omega}    \right]_{\omega +} +\delta(\omega-\omega_0)\left(  \frac{1}{\epsilon^2}+\frac{1}{\epsilon}\left[2\ln\frac{\mu}{\omega_0}  -\frac{5}{2}   \right]  \right),
\end{align}
where the ``$+$"-distribution is definied as
\begin{align}
\int^1_0 du f(u) \Big[g(u,s)\Big]_{u+} 
=&~ \int^1_0 ds \Big[f(u)-f(s)\Big] g(u,s),
\nonumber \\
\int^\infty_0 d\omega f(\omega) \Big[g(\omega,\omega_0)\Big]_{\omega+} 
=&~ \int^\infty_0 d\omega \Big[f(\omega)-f(\omega_0)\Big] g(\omega,\omega_0).
\end{align}

\end{document}